\documentclass[12pt]{article}
\usepackage{graphicx}
\usepackage{amsmath}
\usepackage{amssymb}
\usepackage{bbm}

\begin{document}

\begin{titlepage}

\begin{center}

\begin{flushright}
CERN-PH-TH/2004-169 
%08/09/2004
 %hep-lat/0409xxx
 \\[12ex]
\end{flushright}

\textbf{\large  Finite temperature phase transition \\ in the 4d compact U(1)
lattice gauge theory}
\\[6ex]

{Michele Vettorazzo$^{a,}$\footnote{vettoraz@phys.ethz.ch} and Philippe de
Forcrand$^{a,b,}$\footnote{forcrand@phys.ethz.ch}}
\\[6ex]
{${}^a${\it Institute for Theoretical Physics, ETH Z\"urich,
CH-8093 Z\"urich, Switzerland}\\[1ex]
${}^b${\it CERN, Physics Dept., TH Unit, CH-1211 Geneva 23, Switzerland}}
\\[10ex]
{\small \bf Abstract}\\[2ex]
\begin{minipage}{14cm}{\small We study the phase diagram of the ${\rm 4d}$ compact
${\rm U(1)}$ gauge theory as a function of the number of Euclidean time slices. 
We use the \emph{helicity modulus}
\cite{Vettorazzo:2003fg} as order parameter to probe the phase transitions. The order of
the transition along the phase boundaries is studied and the possibility of a continuum
limit is discussed.
We present new, strong evidence that the $T=0$ bulk phase transition is
first-order.}
\end{minipage}
\end{center}
\vspace{1cm}

\end{titlepage}

%%%%%%%%%%%%%%%%%%%%%%%%%%%%%%%%%%%%%%%%%%%%%%%%%%%%%%%%%%
%INTRODUCTION
%%%%%%%%%%%%%%%%%%%%%%%%%%%%%%%%%%%%%%%%%%%%%%%%%%%%%%%%%%
\section{Introduction}
\label{sec:INTRODUCTION}

The way in which continuum QFT incorporates the definition of `temperature' has many
relations with the notion of \emph{dimensional reduction}: one crucial ingredient is in
fact the \emph{compactification} of the temporal dimension, whose length is tuned as a free
parameter according to 

\begin{equation}
\label{eq:def_of_temperature} T=\frac{1}{\mid \hat L_t \mid}
\end{equation}

\noindent where $T$ is the temperature and $\mid \hat L_t \mid$ is the (dimensionful) time
extension. On the lattice Eq.(\ref{eq:def_of_temperature}) reads

\begin{equation}
T=\frac{1}{L_t a}
\end{equation}

\noindent where $a$ is the lattice spacing and now $L_t$ is a pure integer number; in this
case the notion of finite temperature requires the possibility of taking the limit $a \to
0$, and then to tune $L_t$ accordingly to keep $T$ fixed.

In this paper we consider the ${\rm 4d}$ compact ${\rm U(1)}$ gauge theory on a lattice at
fixed $L_t$. We study its phase diagram and pay particular attention to the points
at which a continuum limit might be extracted, namely the phase boundaries. This is an
interesting generalization of the analogous problem at `zero temperature' ($L_t=\infty$),
still open from the theoretical point of view, where only recently convincing numerical
evidence of the first order nature of the transition was obtained
\cite{Vettorazzo:2003fg,Arnold:2002jk}. Further investigations seem therefore well motivated.\\

The paper is organized as follows. In Sec.~\ref{sec:the_model} we describe the model and
fix the formalism. In Sec.~\ref{sec:the_order_parameter} we introduce the
definition of the order parameter and review some of its properties. In
Sec.~\ref{sec:phase_structure_of_the_model} we study the phase diagram of the model.
Conclusions follow.

%%%%%%%%%%%%%%%%%%%%%%%%%%%%%%%%%%%%%%%%%%%%%%%%%%%%%%%%%%%%%%%%%%%%%%%%
%THE MODEL
%%%%%%%%%%%%%%%%%%%%%%%%%%%%%%%%%%%%%%%%%%%%%%%%%%%%%%%%%%%%%%%%%%%%%%%%
\section{The model}
\label{sec:the_model}

We consider a four dimensional lattice $L_s^3 \cdot L_t$ with periodic boundary conditions.
We associate to each link $U_\mu(r)$ ($r=(x,y,z,t)$) connecting two neighboring sites an
element of the group $U(1)$, namely $e^{i\theta_\mu(r)}$, where $\theta_\mu(r)=ae A_\mu(r)$
is proportional to the vector potential $A_\mu(r)$ and $e$ is the bare gauge coupling. The
system is governed by the Wilson action

\begin{equation}
S=\beta \sum_{r,\mu<\nu} \cos (\theta_P(r))_{\mu\nu}
\end{equation}

\noindent where
$(\theta_P(x))_{\mu\nu}=\theta_\mu(x)+\theta_\nu(x+\hat\mu)-\theta_\mu(x+\hat\nu)-\theta_\nu(x)$
is the so-called plaquette angle, and $\beta=\frac{1}{e^2}$.

%%%%%%%%%%%%CARTOON OF THE TEMPORAL PLAQUETTE%%%%%%%%%%%%%%%%%%%
\begin{figure}[t]
\begin{center}
\includegraphics[height=4.0cm]{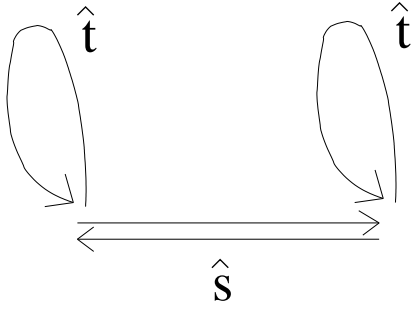}
\end{center}
\caption{\label{fig:temporal_plaquette} Cartoon of a temporal $(s,t)$ plaquette in the case
$L_t=1$. The two spatial links cancel each other, leading to a decoupling of the temporal
links.}
\end{figure}
%%%%%%%%%%%%%%%%%%%%%%%%%%%%%%%%%%%%%%%%%%%%%%%%%%%%%%%%%%%%%%%

The phase structure of this model is well known at zero temperature ($L_t=L_s=L$, $L\to
\infty$): it has a strong coupling confining phase for $\beta < \beta_c$, and a weak
coupling Coulomb phase for $\beta > \beta_c$. There is strong numerical evidence
that the phase transition (at $\beta_c=1.0111331(15)$) is of first order
\cite{Vettorazzo:2003fg}\cite{Arnold:2002jk} . Thus, in the extended phase diagram
$(\beta,1/L_t)$, we have a complete picture of the axis $1/L_t=0$.

There is another limiting case which is  easy to understand, namely $L_t=1$: consider one
temporal plaquette, represented in the cartoon of Fig.~\ref{fig:temporal_plaquette}. In the
case $L_t=1$ the temporal link is already a gauge invariant quantity, a Polyakov loop
itself. Moreover, from Fig.~\ref{fig:temporal_plaquette} it is clear that around a
temporal plaquette the two spatial links are actually the same link considered twice in opposite
orientations, therefore they exactly cancel each other. The temporal plaquette angle is therefore

\begin{equation}
\theta_s(x)+\theta_t(x+\hat s)-\theta_s(x)-\theta_t(x)=\theta_t(x+\hat s)-\theta_t(x)
\end{equation}

\begin{figure}[t]
\begin{center}
\includegraphics[height=6.0cm]{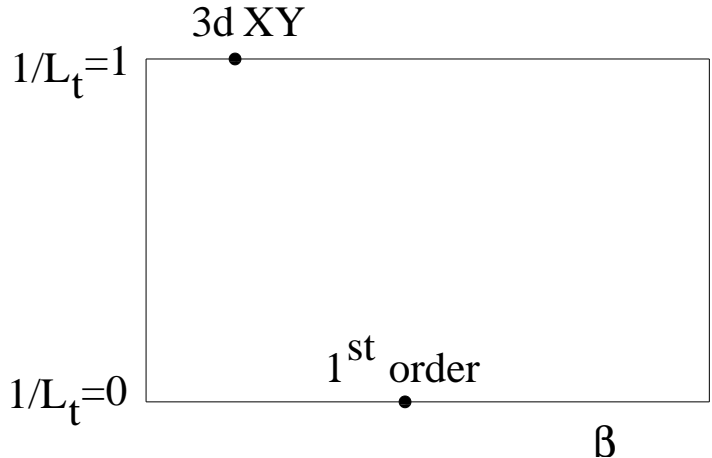}
\end{center}
\caption{\label{fig:phase_diagram_0} Sketch of the phase diagram of the model, collecting
the information about the two limiting situations $L_t=1$ and $L_t=\infty$.}
\end{figure}

The contribution of the temporal plaquettes to the action  is thus

\begin{equation}
\label{eq:temp_action} S_{\rm temp.}=-\beta\sum_{P_{\rm temp.}} \cos
\theta_P=-\beta\sum_{\vec x} \cos (\theta_t(\vec x+\hat s)-\theta_t(\vec x))
\end{equation}

\noindent and it is completely \emph{decoupled} from the spatial plaquettes. Furthermore,
Eq.(\ref{eq:temp_action}) is the action of the ${\rm 3d}$ XY model. One then realizes that in
the case $L_t=1$ the partition function of the model factorizes into

\begin{equation}
Z_{L_t=1}=Z_{\rm 3d \hspace{0.1cm} XY}\cdot Z_{\rm 3d \hspace{0.1cm} LGT}
\end{equation}

\noindent namely we decompose the full partition function into that of the {\rm 3d} XY
model and that of the {\rm 3d} $U(1)$ lattice gauge theory. We know that the former
undergoes a phase transition of second order at $\beta = 0.45420(2)$ \cite{Gottlob:1993jf}, 
while the latter is
always in the confined phase. We summarize the information gained so far about the phase
diagram in Fig.~\ref{fig:phase_diagram_0}.

%%%%%%%%%%%%FIST DETAILS OF THE PHASE STRUCTURE OF THE MODEL%%%%%%%%%%%%%%%%%%%
%%%%%%%%%%%%%%%%%%%%%%%%%%%%%%%%%%%%%%%%%%%%%%%%%%%%%%%%%%%%%%%%%%%%%%%%%%%%%%%

In order to explore the phase structure of the model for a generic value of $L_t$ we need
an order parameter;  we introduce it in the next section.

%%%%%%%%%%%%%%%%%%%%%%%%%%%%%%%%%%%%%%%%%%%%%%%%%%%%%%%%%%%%%%%%%%%%%%%%
%THE ORDER PARAMETER
%%%%%%%%%%%%%%%%%%%%%%%%%%%%%%%%%%%%%%%%%%%%%%%%%%%%%%%%%%%%%%%%%%%%%%%%
\section{The order parameter}
\label{sec:the_order_parameter}

To distinguish between confining and Coulomb behavior, we could monitor the expectation values
of large Wilson loops, or the correlation of two Polyakov loops. However, we can also
construct an observable which directly probes the large-distance properties of the system,
and which has proven itself to be an efficient tool for analyzing a weak first-order transition 
\cite{Vettorazzo:2003fg}: the helicity modulus.

\begin{figure}[t]
\begin{center}
\includegraphics[height=5.0cm]{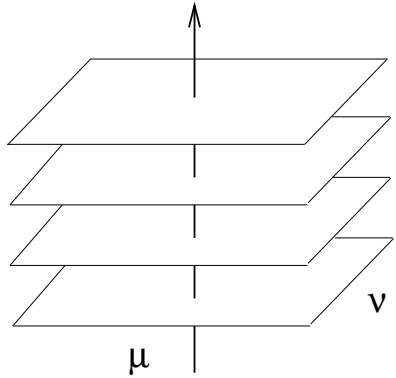}
\end{center}
\caption{\label{fig:spread_flux} Graphical representation of the static external flux
imposed through the $(\mu,\nu)$ orientation.}
\end{figure}

To introduce the helicity modulus for this theory, we first review the notion of
\emph{response function} to an external static field. Consider our four-dimensional lattice
and choose an orientation $(\mu,\nu)$; if we call $L_\rho,L_\sigma$ the two other
(orthogonal) directions, a stack of $L_\rho \cdot L_\sigma$ parallel $(\mu,\nu)$ planes is
thus defined. Let us modify the partition function of the system by imposing an external flux
$\Phi$ through this orientation, as shown in Fig.~\ref{fig:spread_flux}; the  flux is
homogeneously spread over each of the parallel planes, meaning that $\Phi_P=\Phi/L_\mu L_\nu$ is
assigned to each plaquette with orientation $(\mu,\nu)$, as written in the following
equation

%%%%%%%%%%%%HOMOGENEOUSLY SPREAD FLUX%%%%%%%%%%%%%%%%%%%
%%%%%%%%%%%%%%%%%%%%%%%%%%%%%%%%%%%%%%%%%%%%%%%%%%%%%%%%

\begin{equation}
\label{eq:twisted_action} S(\Phi)=-\beta\sum_{P\in (\mu,\nu\hspace{0.1cm}{\rm orient.})}
\cos (\theta_P +\frac{\Phi}{L_\mu L_\nu})-\beta \sum_{P \notin (\mu,\nu \hspace{0.1cm}{\rm
orient.})}\cos \theta_P
\end{equation}

One may wonder if this specific choice of spreading the external flux plays a role in our
construction, but the answer is no, since in this Abelian theory a suitable change of variables
\cite{Vettorazzo:2003fg} can vary at will the distribution of the flux, leaving the
partition function of the system unchanged. Moreover, one can consider another point of
view, and show that the external flux can actually be imposed also via a change of the
boundary conditions from pure periodic to \emph{twisted}: this is very easily accomplished
if we impose, e.g., that each link $U_\mu(r)$ in a particular $\mu$ direction is identified
with its image at distance $L_\nu$ up to a phase factor $e^{i\Phi/L_\mu}$:

\begin{equation}
\theta_\mu(x+L_\nu)=\theta_\mu(x)+\frac{\Phi}{L_\mu}
\end{equation}

Let us now consider the free energy of the system

\begin{equation}
F(\Phi)=-\log Z(\Phi) =-\log \int dU e^{-S(U;\Phi)}
\end{equation}

\noindent and study its behavior in the two phases. In the confining phase
the correlation length $\xi$ is finite (a natural scale in this phase is provided by the string
tension $\sigma=1/\xi^2$), and the effect of the external flux, i.e., of the boundary conditions,
decreases as $e^{-L_\nu/\xi}$. Therefore, in the thermodynamic limit, we obtain the
result

\begin{equation}
\label{eq:free_energy_confined_phase} F(\Phi)={\rm const.}\quad \forall \hspace{0.1cm}\Phi
\end{equation}

The situation is different in the Coulomb phase, where the correlation length of the system
is infinite (the Coulomb potential does not fix any scale in the system) and therefore the
free energy of the system has a non-trivial dependence on $\Phi$. In order to understand
quantitatively this dependence, consider Eq.(\ref{eq:twisted_action}) in the classical
limit $\beta \to \infty$, where all fluctuations are suppressed and the action of
Eq.(\ref{eq:twisted_action}) reduces to the following

\begin{equation}
\label{eq:twisted_action_classical_limit} S_{\rm classic}(\Phi)=-\beta\sum_{P\in
(\mu,\nu\hspace{0.1cm}{\rm orient.})} \cos (\frac{\Phi}{L_\mu L_\nu})+{\rm const.}
\end{equation}

\noindent namely, only the contribution of the twisted plaquettes is present. Consider the
thermodynamic limit at finite temperature, i.e., keeping one size fixed and letting the
others diverge. The product $L_\mu \cdot L_\nu$ diverges, whatever orientation we choose,
therefore it is possible to expand Eq.(\ref{eq:twisted_action_classical_limit}) around its
argument, and obtain

\begin{equation}
\label{eq:classical_flux_dependence} F_{\rm classic}(\Phi)-F_{\rm
classic}(\Phi=0)=\frac{\beta}{2}\Phi^2 \frac{L_\rho L_\sigma}{L_\mu L_\nu}
\end{equation}

\noindent since the sum is over $V=L_\mu L_\nu L_\rho L_\sigma$ identical contributions.
Some comments about this last equation are necessary: first of all, we showed in
\cite{Vettorazzo:2003fg} that for a finite $\beta$, where fluctuations are present,
Eq.(\ref{eq:classical_flux_dependence}) is only slightly modified, namely

\begin{equation}
\label{eq:general_flux_dependence} F_{[{\rm finite} \hspace{0.1cm}\beta]}(\Phi)-F_{[{\rm
finite} \hspace{0.1cm}\beta]}(\Phi=0)=\frac{\beta_R(\beta)}{2}\Phi^2 \frac{L_\rho
L_\sigma}{L_\mu L_\nu}
\end{equation}

\noindent i.e. the bare coupling is replaced by a renormalized coupling, and this is all we
need to effectively take into account the fluctuations of the system. Another remark is
that Eq.(\ref{eq:twisted_action}) is $2\pi$ periodic in $\Phi$, while
Eq.(\ref{eq:general_flux_dependence}) is not. The corresponding modification of
Eq.(\ref{eq:general_flux_dependence}) is exponentially
small in the flux $\Phi$ for small $\Phi$ \cite{Vettorazzo:2003fg}, and plays no role in what follows.\\

Reconsider now Eq.(\ref{eq:general_flux_dependence}), and define

\begin{equation}
\label{eq:helicity_modulus} h(\beta)=\frac{\partial^2 F(\Phi)}{\partial \Phi}\mid_{\Phi=0}
\end{equation}

This quantity is called \emph{helicity modulus} in the literature, and it is a measure of
the \emph{curvature} of the free energy profile around $\Phi=0$. From our previous
discussion about the behavior of $F(\Phi)$ in the two phases, it follows that $h(\beta)$ is
an \emph{order parameter} for our theory. If one performs explicitly the double derivative
w.r.t. $\Phi$, one finds

\begin{equation}
\label{eq:helicity_modulus_spread} h(\beta)=\frac{1}{(L_\mu L_\nu)^2}\lbrace\langle
\sum_{P\in (\mu,\nu \hspace{0.1cm}{\rm orient.})} \beta\cos \theta_P \rangle
-\langle(\sum_{P\in (\mu,\nu \hspace{0.1cm}{\rm orient.})}\beta\sin
\theta_P)^2\rangle\rbrace
\end{equation}

\noindent directly amenable to numerical simulations.

It is important to observe that in our finite temperature system we expect a different
response of the system to fluxes imposed through spatial or temporal orientation.
Consequently the helicity modulus is expected to show different behavior. In the following
we will carefully distinguish the two situations, calling spatial (resp. temporal)
helicity modulus the response to a flux through spatial (resp. temporal) planes.

%%%%%%%%%%%%%%%%%%%%%%%%%%%%%%%%%%%%%%%%%%%%%%%%%%%%%%%%%%%%%%%%%%%%%%%%
%PHASE STRUCTURE OF THE MODEL
%%%%%%%%%%%%%%%%%%%%%%%%%%%%%%%%%%%%%%%%%%%%%%%%%%%%%%%%%%%%%%%%%%%%%%%%
\section{Phase structure of the model}
\label{sec:phase_structure_of_the_model}

%%%%%%%%%%%%CONJECTURED PHASE STRUCTURE OF THE MODEL%%%%%%%%%%%%%%%%%%%
\begin{figure}[t]
\begin{center}
\includegraphics[height=6.0cm]{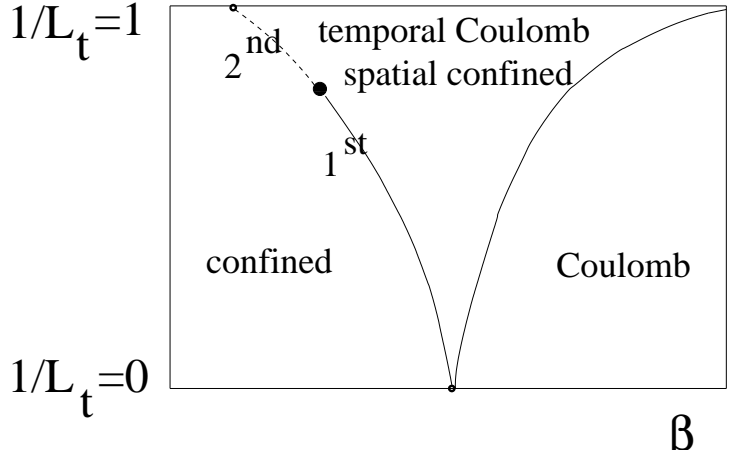}
\end{center}
\caption{\label{fig:phase_diagram_1} Conjectured phase diagram of the model. The phase
boundary on the left separates a confining phase from a partially Coulomb phase, in which
static electric charges propagating in the temporal direction interact Coulombically while
magnetic charges propagating in a spatial direction are confined. The phase boundary on
the right separates this partially Coulomb phase from a pure Coulomb phase. The dotted line
indicates the possibility that the transition remain second order for a range $1 \leq L_t \leq \bar
L_t$ of temporal sizes.}
\end{figure}
%%%%%%%%%%%%%%%%%%%%%%%%%%%%%%%%%%%%%%%%%%%%%%%%%%%%%%%%%%%%%%%%%%%%%%%%%%%%%%%

Let us go back to Fig.~\ref{fig:phase_diagram_0} and try to make an educated guess about
the phase structure of the model. First of all, we expect a phase boundary connecting the
first order transition point at $L_t=\infty$ with the second order point at $L_t=1$ (see
Fig.~\ref{fig:phase_diagram_1}). On the left of this curve we are in a confining phase; on
the right we expect a different behavior of temporal and spatial helicity moduli. A
natural possibility is that the $T=0$ transition between the confined and Coulomb
phases splits for the two orientations at finite temperature. As a consequence we would
have an intermediate phase in which temporal Wilson loops obey perimeter law, while spatial
ones would still show confinement (area law), just like in the high temperature phase of
Yang-Mills theories.

Another important issue is the order of the phase transition along the phase boundary on
the left: we know that in the two limiting cases ($L_t=1$ and $L_t=\infty$) the order of
the transition is different. Two scenarios are possible: either the transition remains more
and more weakly first order, and turns to second order only when $L_t=1$, or there is some
special value $\bar L_t > 1$ for which the transition turns already to second order. This
second scenario is represented in Fig.~\ref{fig:phase_diagram_1} by the dashed line.\\

Let us now consider and analyze separately the two conjectured phase boundaries, to
understand the structure of the model as the thermodynamic limit is approached.

%---------TRANSITION TO COULOMB PHASE--------------------
\subsection{Transition to the Coulomb phase}

In this section we consider the phase boundary on the right in
Fig.~\ref{fig:phase_diagram_1}, the one connecting the temporal Coulomb phase with the pure
Coulomb phase. To explore this transition we want to observe a change of behavior (from area
to perimeter law) in the spatial loops, therefore we have to probe the response of the
system to external fluxes through spatial planes. Equivalently, we will consider the
spatial helicity modulus.

\begin{figure}[th]
\begin{center}
\includegraphics[height=10.0cm,angle=-90]{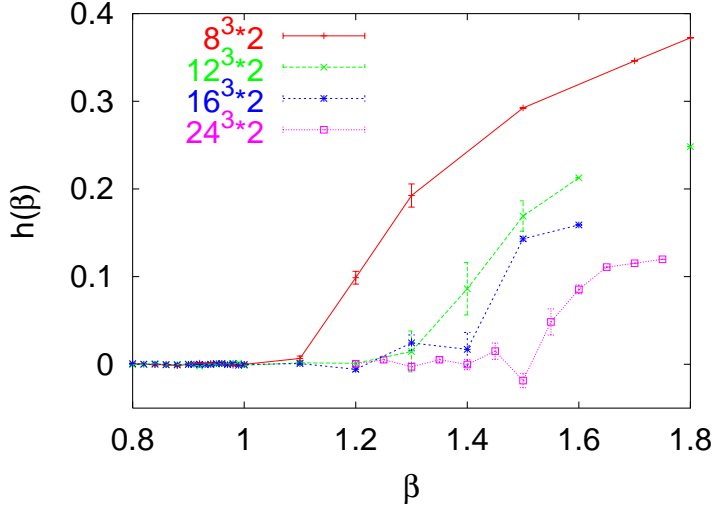}
\end{center}
\caption{\label{fig:hm_stack_SPACE} Spatial helicity modulus as a function of $\beta$,
measured for a temporal size $L_t=2$ and for different $L_s$. The value of $\beta$ at the 
apparent transition point grows with the
spatial  size $L_s$. In the thermodynamic limit this phase boundary disappears.}
\end{figure}

In Fig.~\ref{fig:hm_stack_SPACE} we show the numerical results for the spatial helicity
modulus, measured on a set of lattices with temporal size $L_t=2$ and different spatial sizes $L_s$. 
What is
apparent is that the transition point does not seem to converge at all to some fixed value
$\beta_c$, but moves to higher and higher values with the spatial lattice size. If this
is the case, this transition disappears from the phase diagram in the thermodynamic limit.

%%%%%%%%%%%%PHASE BOUNDARY TO THE COULOMB PHASE%%%%%%%%%%%%%%%%%%%
%%%%%%%%%%%%%%%%%%%%%%%%%%%%%%%%%%%%%%%%%%%%%%%%%%%%%%%%

%%%%%%%%%%%%TEMPORAL FLUX DISTRIBUTION%%%%%%%%%%%%%%%%%%
\begin{figure}[th]
\begin{center}
\includegraphics[height=10.0cm,angle=-90]{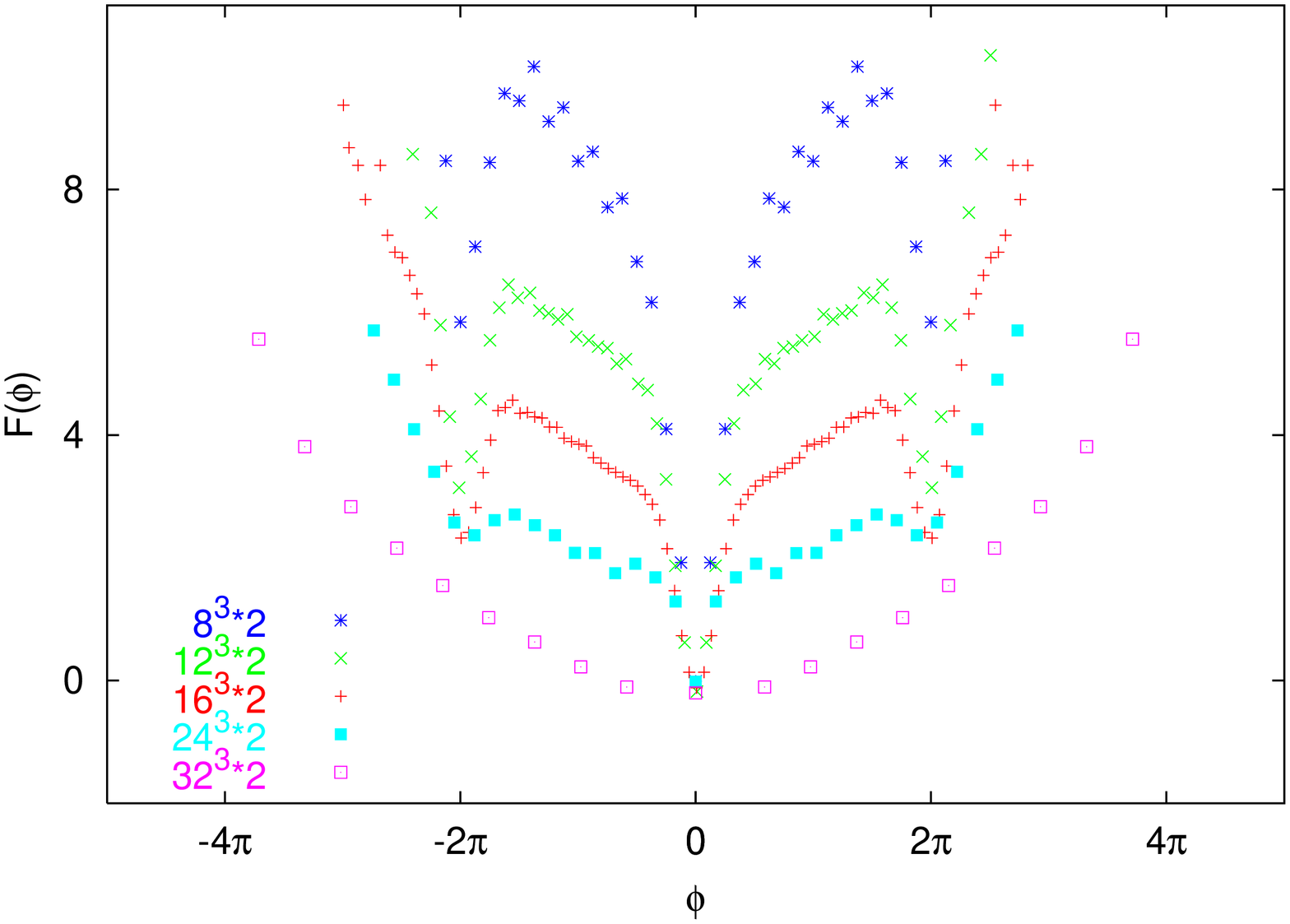}
\end{center}
\caption{\label{fig:flux_free_energy} Flux free energy for lattices with temporal size $L_t=2$ and
increasing spatial size, at $\beta=1.4$. As $L_s\to\infty$ the flux free energy tends to
zero, as expected from Eq.(\ref{eq:free_energy_to_Coulomb_phase}).}
\end{figure}
%%%%%%%%%%%%%%%%%%%%%%%%%%%%%%%%%%%%%%%%%%%%%%%%%%%%%%%%

A first rough way to understand the behavior of the system in this case is given by
Eq.(\ref{eq:general_flux_dependence}). Let us rewrite it here, focusing our attention on
the geometrical factor

\begin{equation}
\label{eq:free_energy_to_Coulomb_phase} 
F_{[{\rm finite} \hspace{0.1cm}\beta]}(\Phi)-F_{[{\rm finite} \hspace{0.1cm}\beta]}(\Phi=0)
=\frac{\beta_R(\beta)}{2}\Phi^2 \frac{L_s L_t}{L_s L_s}
\end{equation}

\noindent where we have highlighted the role of the spatial and temporal directions in this
case. At any fixed value of $\Phi$ and at any finite $\beta$ we get therefore

\begin{equation}
\label{eq:free_energy_to_Coulomb_phase_2} F(\Phi) - F(0) \sim \frac{\beta_R}{L_s}\to 0
\hspace{0.25cm} {\rm as} \hspace{0.25cm}  L_s \to \infty
\end{equation}

\noindent meaning that the flux free energy always vanishes in the thermodynamic
limit; this prediction is nicely confirmed in Fig.~\ref{fig:flux_free_energy}, where
$F(\Phi)$ is measured from the flux distribution $\nu(\Phi)=e^{-F(\Phi)}$ for different
spatial extensions $L_s$. As a consequence, no phase transition is present between the
mentioned phases for any finite value of $L_t$, but just the interplay of the coupling and
of the spatial lattice size mimics
a transition.\\

An alternative way to understand this finite size effect is the following.  At any given 
value of $\beta$ there is a finite density of magnetic monopole currents. In
particular, a wrapping (non-contractible) time-like current loop, far apart from its companion
anti-loop, can disorder all the spatial planes in the lattice, or, equivalently, a spatial
Wilson loop of arbitrary size; this, in turn, produces confinement
\cite{Vettorazzo:2003fg}. Call $\rho(\beta)$ the density of currents at a given value of
$\beta$; we know \cite{DeGrand:1980eq} that

\begin{equation}
\rho(\beta)\sim e^{-c\beta}
\end{equation}

\noindent where $c$ is a constant, for $\beta \gg 1$. Let $n(l)dl$ be the distribution
function indicating the fraction of monopole loops of length between $l$ and $l+dl$. The
transition occurs approximately when each configuration contains a pair of non-contractible 
monopole loops, namely when

\begin{equation}
\label{eq:phase_trans_condition_0} \rho(\beta) L_s^3 \int_{l \geq 2 L_t} n(l)dl \gtrsim 1
\end{equation}

\noindent where the integration domain $l \geq 2  L_t$ represents at least a necessary
condition to have wrapping loops; on average we expect it to be also sufficient. The last
piece of information we need is

\begin{equation}
\frac{dn(l)}{dl}\sim e^{-\delta l}
\end{equation}

\noindent motivated by the fact that the free energy cost of a current is proportional to
its length $l$ (notice that this is true only in the Coulomb phase, where the coefficient
$\delta > 0$). Let us put together all the ingredients and rewrite
Eq.(\ref{eq:phase_trans_condition_0})

\begin{equation}
\label{eq:phase_trans_condition_1}  L_s^3 e^{-c\beta} \int_{2 L_t}^\infty e^{-\delta l}dl =
L_s^3 e^{-c\beta}\frac{e^{-2 \delta L_t}}{\delta}\gtrsim 1
\end{equation}

\noindent which contains all the information we need. At fixed volume, the inequality is
satisfied only for sufficiently low $\beta$; when $\beta$ is large enough we are always in
an ordered phase, as indicated by our numerical finding (spatial helicity modulus different
from zero). In the limit $L_s\to\infty$ at fixed $L_t$, this equation is always satisfied,
indicating that the system is always disordered. It is only in the limit $L_t \to \infty$
(because of the term $e^{-2 \delta L_t}$) that we can obtain an ordered phase.
Interestingly, this equation predicts a pseudocritical coupling

\begin{equation}
\label{eq:critical_coupling_vs_L} \beta_c \sim \log L_s
\end{equation}

\noindent contrary to the simpler argument in Eq.(\ref{eq:free_energy_to_Coulomb_phase_2})
which gives

\begin{equation}
\beta_c \sim  L_s
\end{equation}

\noindent
The first argument is completely classical, therefore we expect that
Eq.(\ref{eq:critical_coupling_vs_L}) describes the physics better.

%---------TRANSITION TO CONFINED PHASE--------------------

%%%%%%%%%%%%PHASE BOUNDARY TO THE CONFINED PHASE%%%%%%%%%%%%%%%%%%%
\begin{figure}[th]
\begin{center}
\includegraphics[height=10.0cm,angle=-90]{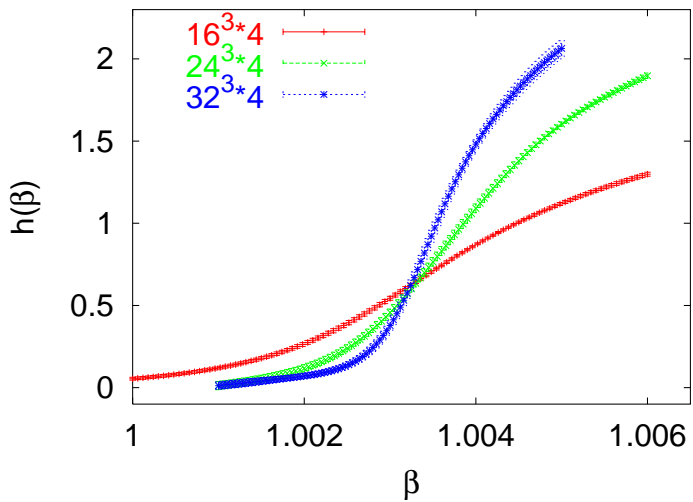}
\end{center}
\caption{\label{fig:hm_stack_TIME} Temporal helicity modulus as a function of $\beta$,
measured on systems of temporal size $L_t=4$ and different spatial sizes $L_s$.}
\end{figure}
%%%%%%%%%%%%%%%%%%%%%%%%%%%%%%%%%%%%%%%%%%%%%%%%%%%%%%%%

%%%%%%%%%%%%FSS SCALING ANALYSIS%%%%%%%%%%%%%%%%%
\begin{figure}[th]
\hspace*{-0.7cm}
\includegraphics[height=8.cm,angle=-90.]{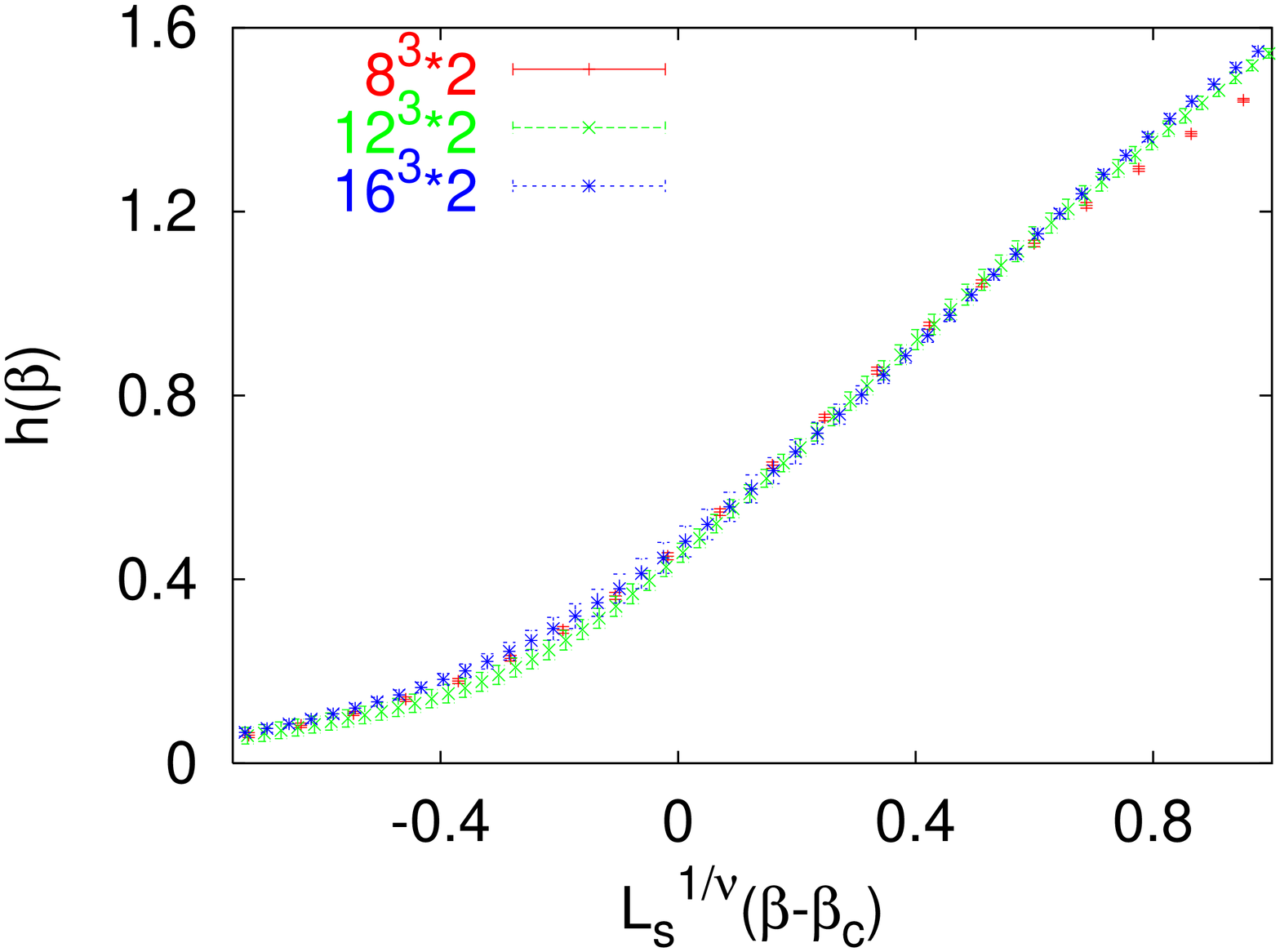}
\includegraphics[height=8.cm,angle=-90.]{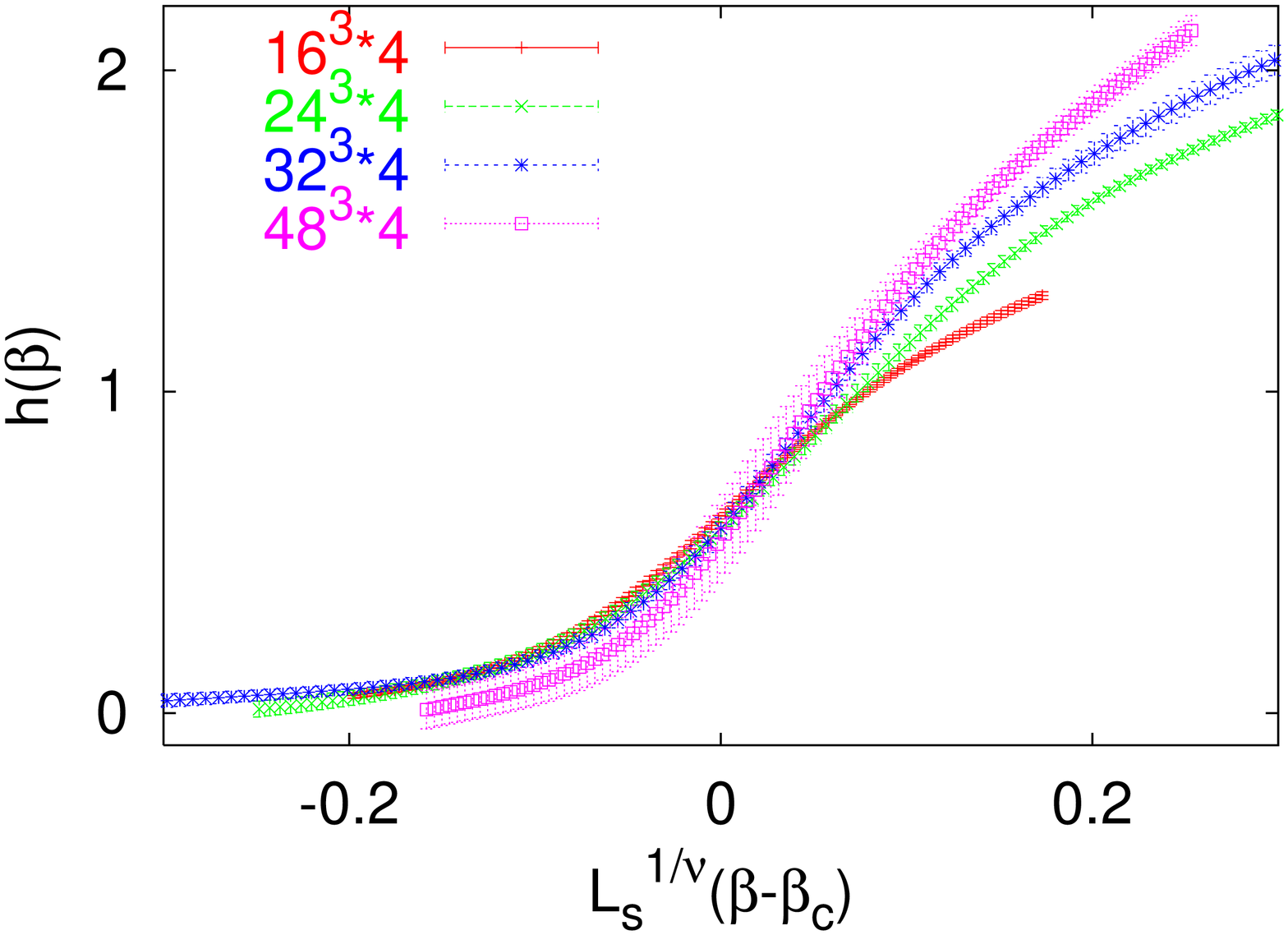}
\caption{\label{fig:FSS_analysis_of_the_data} Finite-size scaling analysis of the temporal helicity
modulus for lattices of temporal extent $L_t=2$ (left) and $L_t=4$ (right). The scaling
window becomes narrower as $L_t$ increases, consistent with the approach to a first order
transition.}
\end{figure}
%%%%%%%%%%%%%%%%%%%%%%%%%%%%%%%%%%%%%%%%%%%%%%%%%%%%%%%%%%%%%%%%%%%%

\subsection{Transition to the confining phase}
\label{sec:transition_to_the_confining_phase}

\begin{center}
\begin{figure}[th]
\hspace*{-0.7cm}
\begin{minipage}{4.9cm}
\includegraphics[height=5.5cm,angle=-90.]{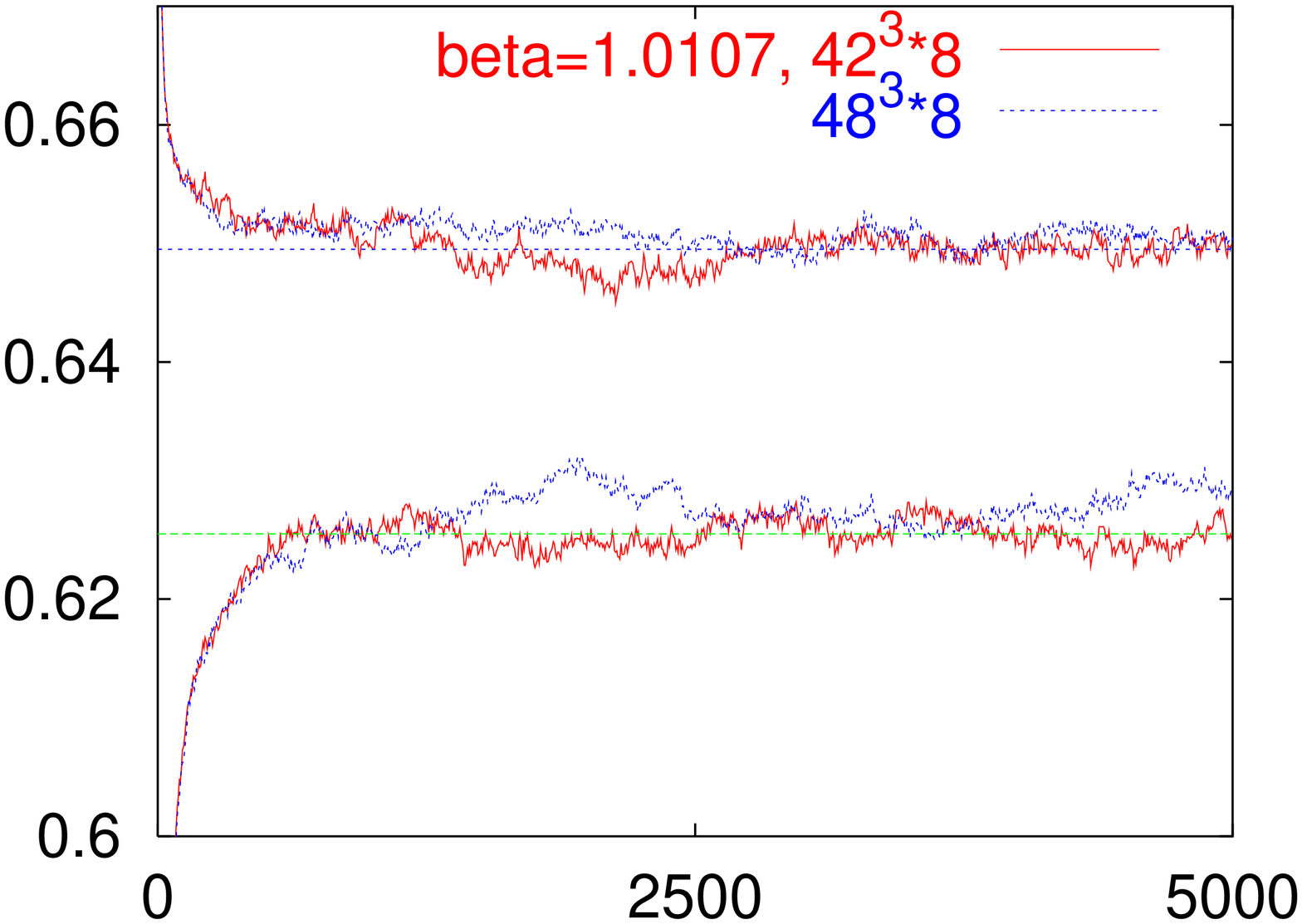}
\end{minipage}
\begin{minipage}{4.9cm}
\includegraphics[height=5.5cm,angle=-90.]{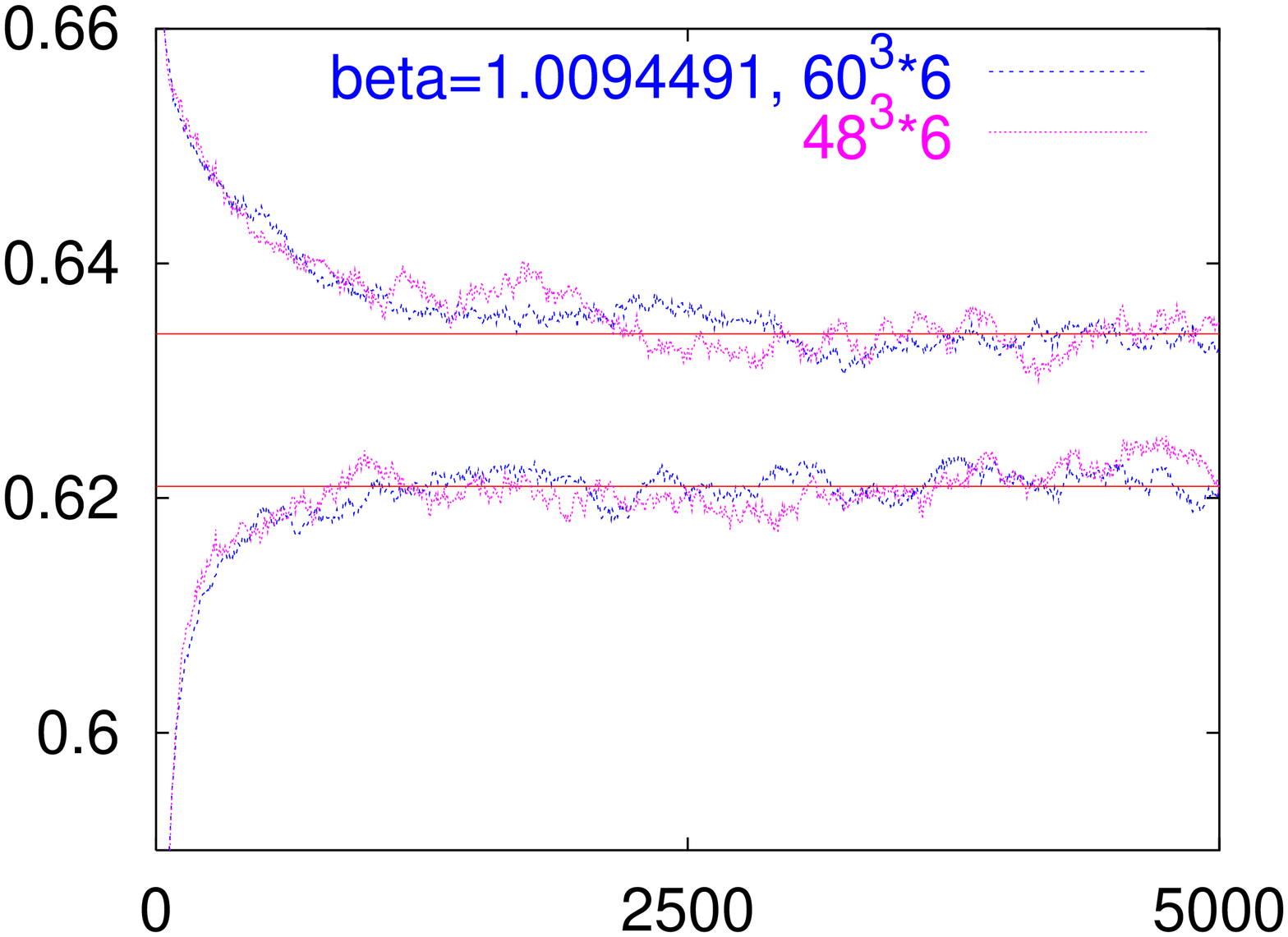}
\end{minipage}
\begin{minipage}{4.9cm}
\includegraphics[height=5.5cm,angle=-90.]{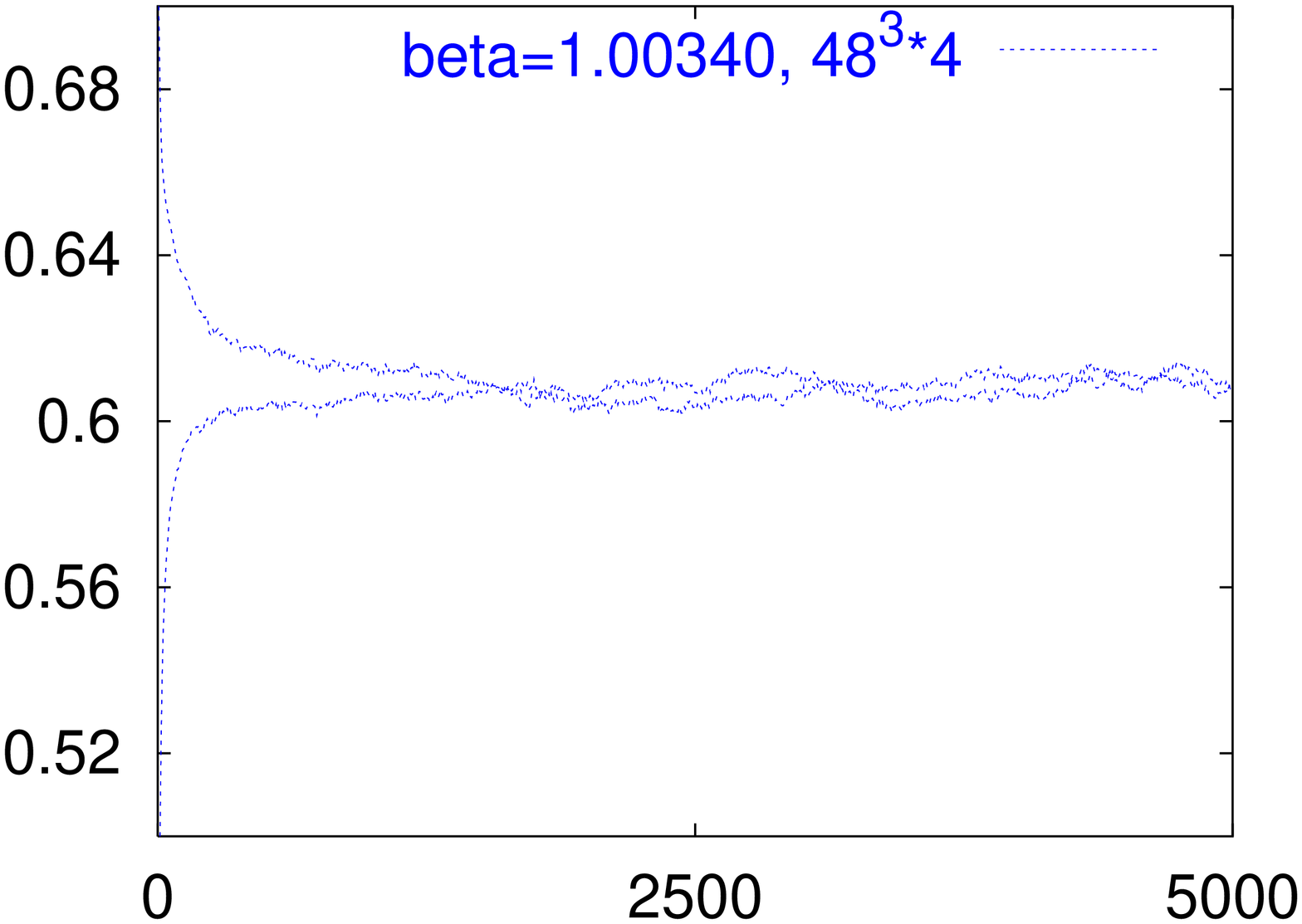}
\end{minipage}
\caption{\label{fig:measure_latent_heat} Time history of the plaquette, starting from an ordered 
(upper curves) and disordered (lower curves) initial state. 
The latent heat is defined as the difference of the values of the
two plateaus. From left to right the cases $L_t=8,6,4$ are studied. For $L_t=8,6$, where
the latent heat is certainly different from zero, we checked that the thermodynamic limit
had been reached by comparing different spatial volumes.}
\end{figure}
\end{center}

\begin{figure}[th]
\begin{center}
\includegraphics[height=10.0cm,angle=-90]{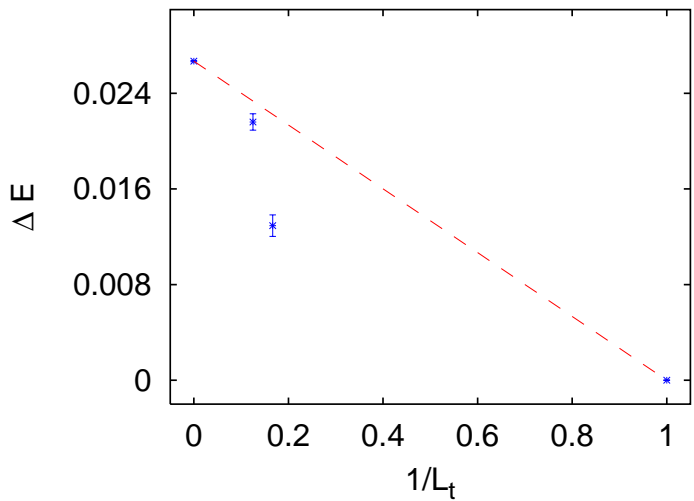}
\end{center}
\caption{\label{fig:latent_heat_vs_Nt} Latent heat $\Delta E$ of the transition versus
$1/L_t$. The displayed data correspond to $L_t=\infty$, then $L_t=8,6,1$. The cases $L_t=4$
and $2$ are not displayed because the question whether the latent heat is already zero
cannot be resolved with our numerical measurements. The dashed line corresponds to the
simplest, linear ansatz.}
\end{figure}

Let us now turn our attention to the left phase boundary in Fig.~\ref{fig:phase_diagram_1},
and to the temporal helicity modulus. Eq.(\ref{eq:general_flux_dependence}) gives

\begin{equation}
\label{eq:free_energy_to_confined_phase} F_{[{\rm finite}
\hspace{0.1cm}\beta]}(\Phi)-F_{[{\rm finite}
\hspace{0.1cm}\beta]}(\Phi=0)=\frac{\beta_R(\beta)}{2}\Phi^2 \frac{L_s L_s}{L_s L_t}\sim L_s \beta_R(\beta)
\end{equation}

\noindent therefore in the thermodynamic limit, at the transition point, the signal is
infinitely enhanced; the transition point is determined (up to finite size effects) by the
behavior of $\beta_R(\beta)$, which is related to the free energy of a monopole current, an
order parameter by itself. In Fig.~\ref{fig:hm_stack_TIME} we show one set of results for
$L_t=4$ and three spatial volumes (the data are reweighted \emph{$\grave{a}$ la}
Ferrenberg-Swendsen \cite{Ferrenberg:xy}). From a finite size scaling analysis of such data
we can extract the numerical value of $\beta_c$ as a function of the time extension $L_t$,
and the \emph{order} of the transition. For small values of $L_t$ we want to decide if the
transition is first or second order. Our strategy is to assume the second order scenario
(the universality class is then that of the ${\rm 3d~XY}$ model, as for $L_t=1$) and then
check the quality of the data collapse. We thus fix $\nu=0.6723$ (the value of the ${\rm
3d}$ XY model \cite{Hasenbusch:1999cc}) and rescale the $\beta$ axis according to

\begin{equation}
\beta \to L_s^{1/\nu}(\beta-\beta_c)\sim (\frac{L_s}{\xi})^{1/\nu}
\end{equation}

\noindent where we have to tune $\beta_c$ only. The results of such analysis for the
lattices with $L_t=2,4$ are presented in Fig.~\ref{fig:FSS_analysis_of_the_data}. Comparing
the excellent collapse of the curves in the case $L_t=2$ with the mediocre collapse in the
case $L_t=4$, one gets a first qualitative indication of a second order transition turning
into a first order one. Moreover, one can speculate that for small values of $L_t > 1$ the
transition still is second order, as indicated in the cartoon of the phase diagram of
Fig.~\ref{fig:phase_diagram_1}. Since this issue is relevant to the possibility of
extracting a continuum limit, we further investigated this problem directly by measuring the
\emph{latent heat} of the transition as a function of $L_t$. We took the direct approach of
monitoring the Monte-Carlo history of the plaquette, starting from hot and cold 
(disordered and ordered) initial
configurations. When a metastability is observed, we measure the distance between the two
plateau values. We repeated the measurements  for different volumes in order to keep the
finite volumes effects under control. The plaquette histories are displayed in 
Fig.~\ref{fig:measure_latent_heat}, and the estimated latent heats $\Delta E(L_t)$ in 
Fig.~\ref{fig:latent_heat_vs_Nt}. Table I summarizes our measurements of the critical parameters. 
As far as we know there are no theoretical predictions
about the functional form that we observe for $\Delta E(L_t)$. The data anyway seem to suggest 
a scenario in which the latent heat drops to zero already for $\bar L_t \sim 4$,
at $\beta_c(L_t) \sim 1$. 

%%%%%%%%%%%%LATENT HEAT: HOW DO WE MEASURE IT%%%%%%%%%%%%%%%%%
%%%%%%%%%%%%%%%%%%%%%%%%%%%%%%%%%%%%%%%%%%%%%%%%%%%%%%%%%%%%%%%%%%%%

%%%%%%%%%%%%LATENT HEAT VERSUS L_t%%%%%%%%%%%%%%%%%%%%%%
%%%%%%%%%%%%%%%%%%%%%%%%%%%%%%%%%%%%%%%%%%%%%%%%%%%%%%%%

%%%%%%%%%%%%%%%%%%%%%%%%%%%%%%%%%%%%%%%%%%%%%%%%%%%%%%%%%%%%%%%%%%%%%%%%%%%
%TABLE: LATENT HEAT VERSUS TEMPERATURE
%%%%%%%%%%%%%%%%%%%%%%%%%%%%%%%%%%%%%%%%%%%%%%%%%%%%%%%%%%%%%%%%%%%%%%%%%%%
\begin{table}
\begin{center}
\begin{tabular}{|c|c|c|}\hline

$L_t$  & $\Delta E$ & $\beta_c$ \\

\hline

$\infty$ &   0.02672(6) \cite{Arnold:2002jk} &  1.011331(15) \cite{Arnold:2002jk} \\ \hline

8 &  0.0216(7) & 1.0107(1) \\ \hline

6 &  0.0129(9) & 1.009449(1) \\ \hline

4 &  - & 1.00340(1) \\ \hline

3 &  -  & 0.989(1) \\ \hline

2 &  -  & 0.9008(3) \\ \hline

1 &  0  & 0.45420(2) \cite{Gottlob:1993jf} \\ \hline

\end{tabular}
\caption{\label{tab:delta_E_vs_Lt} Latent heat $\Delta E$ and transition coupling $\beta_c$ 
versus $L_t$. $\Delta E$ decreases very quickly for $L_t < 6$, opening the possibility of
a second order phase transition for $L_t\lesssim 4$.}
\end{center}
\end{table}

We cannot exclude a small but non-zero latent heat of course. However, 
Eq.(\ref{eq:free_energy_to_confined_phase})
suggests that the transition turns from first- to second-order when $\beta_R(\beta_c)$ 
vanishes as the transition point $\beta_c$ is approached from above, since then the
magnitude of the jump in the helicity modulus at the phase transition also vanishes.
The rapid decrease of $\beta_R$ with $\beta$ observed at zero temperature in \cite{Vettorazzo:2003fg}
makes it unlikely that $\beta_R$ would remain non-zero all the way down to 
$\beta_c(L_t=1) \approx 0.45420$. Thus, we consider the scenario where the transition
is second-order for a range of temporal sizes $[1,\bar L_t]$ as a likely alternative.

This opens the intriguing possibility of a continuum limit, obtained by approaching 
any point along such a second-order line. Note however that the number $L_t$ of time-slices 
cannot exceed $\bar L_t$, so that one cannot strictly speaking talk about the $U(1)$ system 
being at ``finite temperature''. Rather, we have a system of $L_t \leq \bar L_t$ coupled
${\rm 3d-}$systems. Moreover, as one approaches the critical point, only the 
temporal correlation length $\xi_t/a$ diverges in lattice units, whereas the spatial correlation
length $\xi_s/a$ remains finite. Thus, keeping the ``physical'' correlation length $\xi_t$ constant
would imply not only $a \to 0$, but also $\xi_s \to 0$, so that the spatial planes would be 
\emph{completely disordered}. The physical relevance of this situation seems marginal.

%%%%%%%%%%%%%%%%%%%beta_c VERSUS L_t%%%%%%%%%%%%%%%%%%%%
\begin{figure}[th]
\begin{center}
\includegraphics[height=10.0cm,angle=-90]{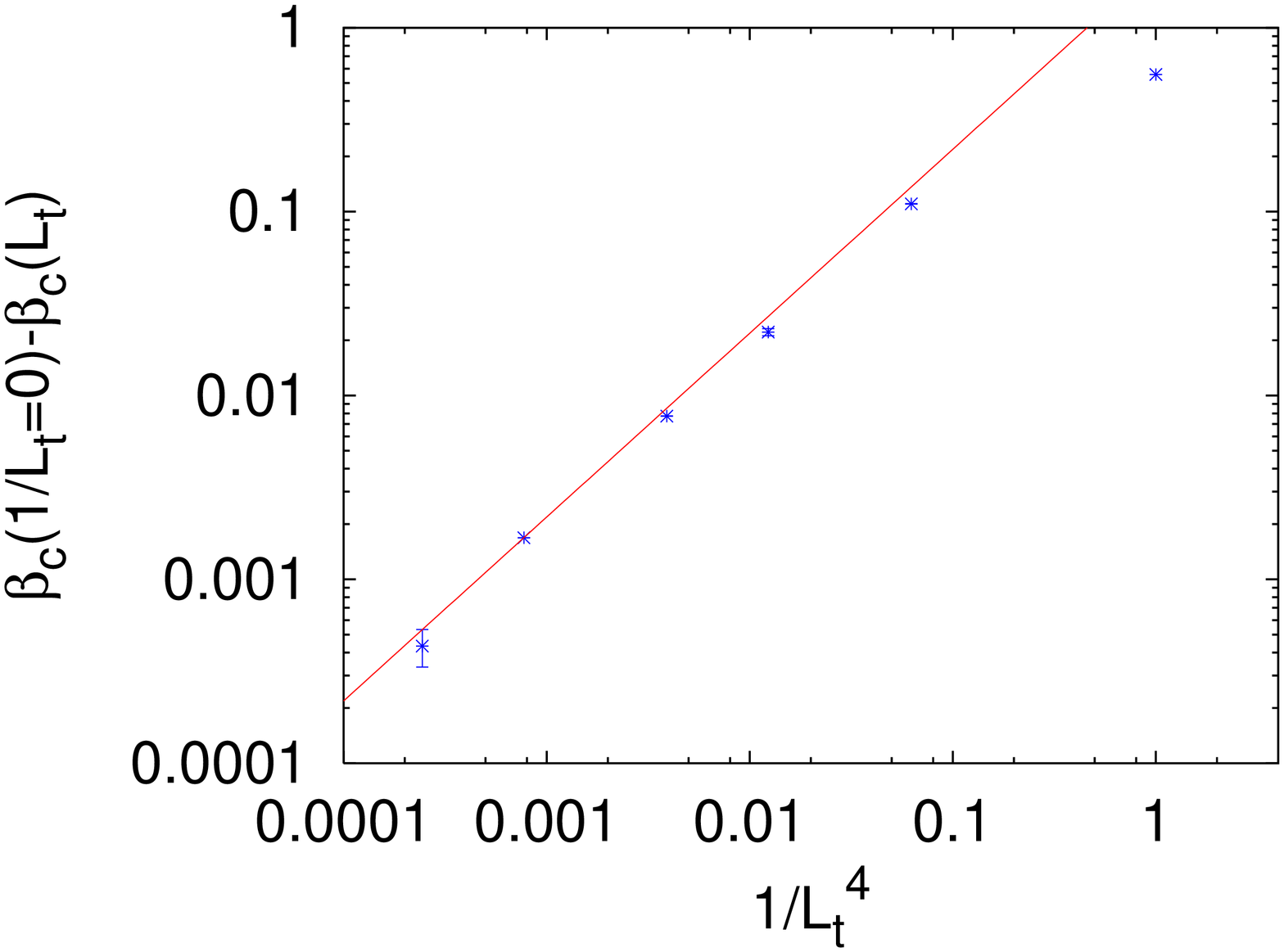}
\end{center}
\caption{\label{fig:Nt} Shift of $\beta_c$ as a function of the Euclidean time extent $L_t$. 
The scaling ansatz shown is
$\frac{c}{L_t^{1/\nu}}$, with $c=2.18$ and $\nu=1/4$. An exponent
$\nu = 1/d$ is the typical signature of a first-order phase transition.}
\end{figure}
%%%%%%%%%%%%%%%%%%%%%%%%%%%%%%%%%%%%%%%%%%%%%%%%%%%%%%%%

The data in Table I can also be used for another kind of analysis,
namely to study the dependence of $\beta_c$ on $L_t$ as $L_t\to
\infty$. The coupling at the transition should in fact approach its asymptotic value
$\beta_c(1/L_t=0)$ with typical finite-size corrections:

\begin{equation}
\beta_c(L_t) \approx \beta_c(1/L_t=0) + \frac{c}{L_t^{1/\nu}}
\end{equation}

\noindent where $c$ is a constant, and $\nu=\frac{1}{d}=\frac{1}{4}$ for 
a first-order transition. This is indeed the case, as shown in
Fig.~\ref{fig:Nt}, where the straight line corresponds to $1/L_t^4$ corrections.
This represents a qualitatively new and strong evidence of the first
order nature of the transition in the zero temperature limit.

%%%%%%%%%%%%%%%%%%%%%%%%%%%%%%%%%%%%%%%%%%%%%%%%%%%%%%%%%%%%%%%%%%%%%%%%
%CONCLUSIONS
%%%%%%%%%%%%%%%%%%%%%%%%%%%%%%%%%%%%%%%%%%%%%%%%%%%%%%%%%%%%%%%%%%%%%%%%
\section{Conclusions}

%%%%%%%%%%%%PHASE DIAGRAM OF THE MODEL%%%%%%%%%%%%%%%%%
\begin{figure}[th]
\begin{center}
\includegraphics[height=10.cm,angle=-90.]{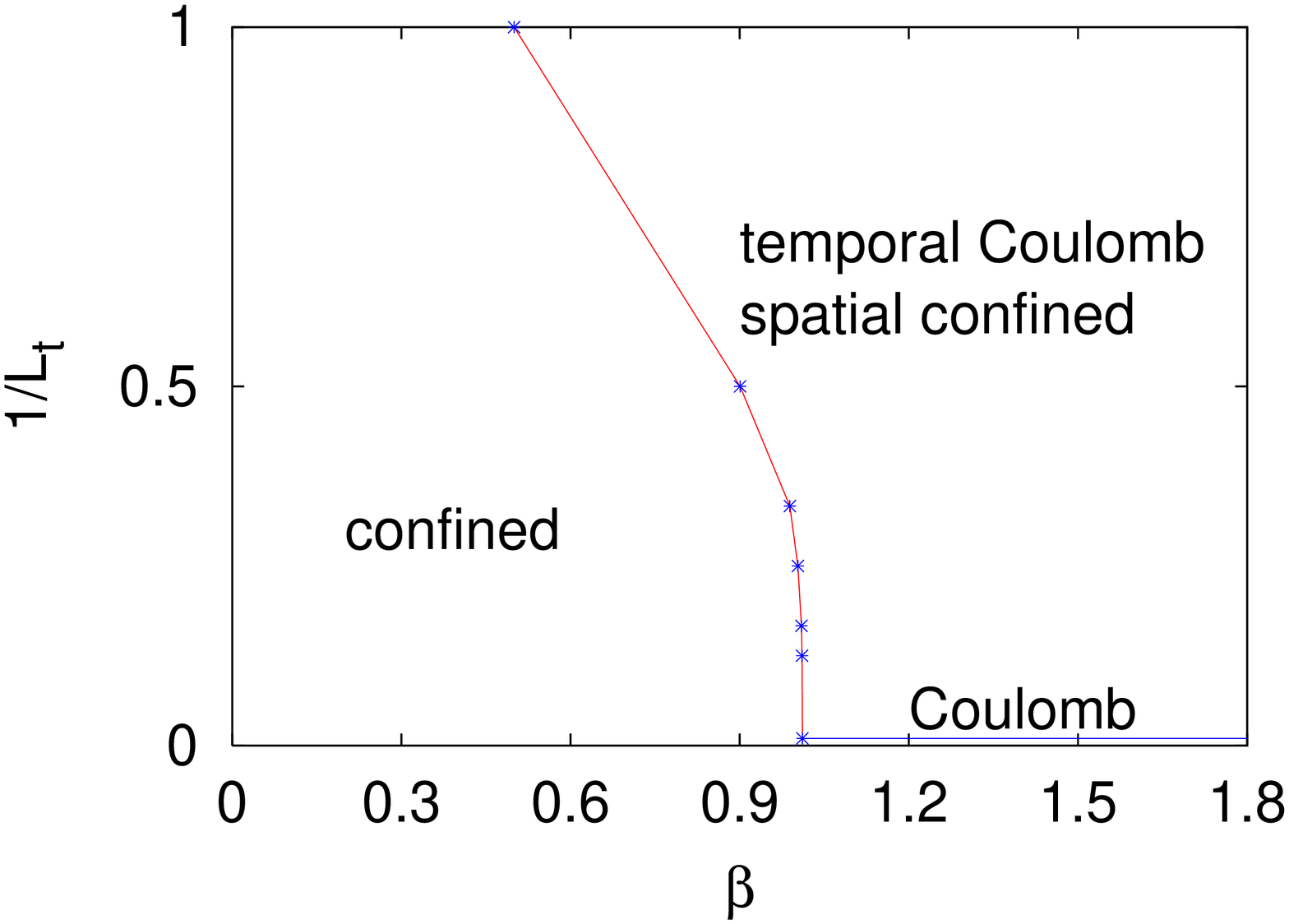}
\end{center}
\caption{\label{fig:final_phase_diagram} Phase diagram of the model. The phase boundary is
obtained from a finite-size scaling analysis of the temporal helicity modulus. Three phases are
indicated: a confining phase on the left of the phase boundary, a temporal Coulomb phase on
the right of the boundary at finite $L_t$, and the pure Coulomb phase on the right of the
boundary and for $L_t=\infty$.}
\end{figure}
%%%%%%%%%%%%%%%%%%%%%%%%%%%%%%%%%%%%%%%%%%%%%%%%%%%%%%%%%%%%%%%%%%%%

In this paper we have investigated the phase diagram of the $4d$ $U(1)$ lattice gauge
theory in the extended plane $(\beta,1/L_t)$, where $L_t$ is the number of time-slices
of the system. A summary of our findings is provided by
Fig.~\ref{fig:final_phase_diagram} where the phase diagram of the model is shown. Only one
phase boundary is present, whose position we have determined. It divides a confining phase
on the left and a temporal Coulomb phase (with broken global $U(1)$ symmetry) on the right for any finite $L_t$; the pure
Coulomb phase exists  only for $L_t=\infty$ and $\beta>\beta_c=1.0111331(15)$. This phase diagram is
remarkably similar to what we find in Yang-Mills theories, where spatial confinement
persists at all temperatures.

The order parameter we used, the helicity modulus, reduces naturally, via dimensional
reduction, to the ordinary helicity modulus known in condensed matter. We can therefore say 
that our order parameter is a generalization for gauge theories of the already known quantity.

The transition for any finite value of $L_t$ is first order, except perhaps for
$L_t \lesssim 4$. In this case the transition becomes so weak that we cannot determine its
order, but consider likely the possibility of a second order phase transition. 
This question could be studied by more powerful numerical methods. For instance,
one could implement spatial $C$-periodic boundary conditions, which reduce the global
$U(1)$ symmetry to $Z_2$, then measure the height of the free energy barrier between
the two vacua as a function of the system size.

However, even if the transition is confirmed to be second order for $L_t \lesssim 4$,
the relevance of this result 
to a continuum limit of the finite temperature $U(1)$ theory is marginal: 
in order to introduce the notion of temperature

\begin{equation}
T=\frac{1}{L_t a}
\end{equation}

\noindent one must tune $L_t$ as a function of $a\to 0$ in order to maintain a finite $T$. 
If, as in our case,
$L_t$ cannot be larger than a certain $\bar L_t$, the continuum limit cannot be taken,
and we should not really talk about ``finite temperature''.

Finally, we would like to mention the old paper \cite{Kronfeld:1987ri}, where the authors
considered the Georgi-Glashow model with gauge group $SU(2)$ at finite temperature; this model
reduces to our $U(1)$ model in the limit of complete gauge symmetry breaking (where $U(1)$
is the symmetry remnant of the broken $SU(2)$). The conjecture of the authors about the
phase diagram of our model, contained in their Fig.~$1$, is completely consistent with
our findings.

An even older, correct, phase diagram can be found in \cite{Svetitsky:1985ye}. 
The persistence of the spatial area law at finite temperature was first proven 
in \cite{Borgs:1985qh}.

%%%%%%%%%%%%%%%%%%%%%%%%%%%%%%%%%%%%%%%%%%%%%%%%%%%%%%%%%%%%%%%%%%%%%%%%
%ACKNOWLEDGMENT
%%%%%%%%%%%%%%%%%%%%%%%%%%%%%%%%%%%%%%%%%%%%%%%%%%%%%%%%%%%%%%%%%%%%%%%%
\section{Acknowledgement}
We gratefully acknowledge J\"urg Fr\"ohlich, Oliver Jahn and Ben Svetitsky for useful discussions.

%%%%%%%%%%%%%%%%%%%%%%%%%%%%%%%%%%%%%%%%%%%%%%%%%%%%%%%%%%%%%%%%%%%%%%%%
%BIBLIOGRAPHY
%%%%%%%%%%%%%%%%%%%%%%%%%%%%%%%%%%%%%%%%%%%%%%%%%%%%%%%%%%%%%%%%%%%%%%%%

\end{document}